\documentclass{article}

\usepackage{amsfonts,natbib}
\usepackage{amsmath, graphicx, epsfig,psfrag, verbatim}

\usepackage{epstopdf}

\newcommand{\bmu}{\mbox{\boldmath{$\mu$}}}
\newcommand{\bthet}{\mbox{\boldmath{$\theta$}}}
\newcommand{\balph}{\mbox{\boldmath{$\alpha$}}}
\newcommand{\bomega}{\mbox{\boldmath{$\omega$}}}
\newcommand{\bbet}{\mbox{\boldmath{$\beta$}}}
\newcommand{\bzet}{\mbox{\boldmath{$\zeta$}}}
\newcommand{\bvthet}{\mbox{\boldmath{$\vartheta$}}}

\numberwithin{equation}{section}

\begin{document}
\baselineskip20truept \pagestyle{plain}

\title{Simulation based sequential Monte Carlo methods for discretely observed Markov processes}

\author{Peter Neal}

\maketitle

\begin{abstract}
Parameter estimation for discretely observed Markov processes is a
challenging problem. However, simulation of Markov processes is
straightforward using the Gillespie algorithm. We exploit this ease
of simulation to develop an effective sequential Monte Carlo (SMC)
algorithm for obtaining samples from the posterior distribution of
the parameters. In particular, we introduce two key innovations,
coupled simulations, which allow us to study multiple parameter
values on the basis of a single simulation, and a simple, yet
effective, importance sampling scheme for steering simulations
towards the observed data. These innovations substantially improve
the efficiency of the SMC algorithm with minimal effect on the speed
of the simulation process. The SMC algorithm is successfully applied
to two examples, a Lotka-Volterra model and a Repressilator model.
\end{abstract}


\section{Introduction} \label{S:Intro}

Markov processes are used to model a wide range of biological
systems, for example, epidemic models (\cite{Bailey}), predator-prey
models (\cite{BWK}) and gene regulatory systems (\cite{Toni09}). The
above are examples of individual-based compartmental models, where
the system spends an exponentially distributed length of time in the
current state before making a transition to a new state. Both the
mean length of stay in the current state and the probability of
transition to a particular new state are only dependent upon the
current state of the system and the model parameters.

Parameter estimation for Markov processes is straightforward if the
entire continuous time process is observed. However, this is rarely
the case with the system often observed at a discrete collection of
points with either complete or partial observations of the system
occurring at the observation points. Observation of the process at a
discrete set of points does not yield a tractable likelihood for
parameter estimation. One solution in a Bayesian context is to use
data augmentation MCMC (Markov chain Monte Carlo), see for example
\cite{BWK}. However, the data augmentation MCMC algorithm will often
require large scale data imputation since typically many events,
possibly running into the hundreds or thousands, will have taken
place between each pair of observations. This can often result in
poor mixing of the MCMC algorithm given the strong correlation
between the model parameters and the imputed data.

It is necessary to consider alternatives to data augmentation MCMC.
Assuming that the number of individuals in each of the compartments
is relatively large, the Markov process can be approximated by a
system of ordinary differential equations (ode), \cite{Kurtz70} with
a diffusive limit about the ode solution, \cite{Kurtz71}. This has
been exploited to create diffusion and linear noise approximation
algorithms, see for example \cite{GW11} and \cite{FGS}. However,
there are many biological systems, including the gene regulatory
system considered in Section \ref{S:Rep}, where the total number in
a component can often be very small, even 0 and the above
approximations are often not appropriate.

An alternative to MCMC which has been applied to Markov processes is
Approximate Bayesian Computation (ABC), see \cite{Tav} and
\cite{BZB}. ABC is a simulation based method where in its simplest
form data, $\mathbf{x}$, is simulated using the model for a given
set of parameters $\bthet$ chosen from the prior distribution on
$\bthet$. If $\mathbf{x}$ is sufficiently {\it close} to the
observed data, $\mathbf{x}^\ast$, then the parameters $\bthet$ are
accepted from the (approximate) posterior distribution of $\bthet |
\mathbf{x}^\ast$. (If {\it sufficiently close to} is replaced by
{\it equal to} the accepted values are independent and identically
distributed observations from the posterior distribution of
$\bthet$.) Using ABC for Markov processes is straightforward given
the ease with which Markov processes can be simulated using the
Gillespie algorithm \cite{Gill}. However, as noted in \cite{WKP},
the probability of simulating the entire Markov process
 and getting $\mathbf{x}$ close to $\mathbf{x}^\ast$ is
(extremely) small. The solution proposed in \cite{WKP} is, in the
case where the Markov process is completely observed at each
observation point, to exploit the Markov structure of the process
and consider each interval (between observation points) separately.
In particular, \cite{WKP} factorise the likelihood and estimate the
posterior distribution of the parameters based upon each interval
(component in the factorised likelihood) before combining the
estimates to gain an overall estimate of the posterior distribution.

In this paper, we take the approach of \cite{WKP} piecewise
simulation of the Markov process as our starting point. More
precisely, we outline a sequential Monte Carlo (SMC) algorithm which
is applicable in the case where the Markov process is only partially
observed at each observation time point. The SMC algorithm updates
the posterior distribution of the parameters $\bthet$ after each
time point and uses simulation between successive observation points
with parameters drawn from the current posterior distribution to
update the posterior distribution. The sequential evaluation of the
posterior distribution, improves upon simply choosing parameters
from the prior, which is the case in \cite{WKP}. The basic SMC
algorithm is described in Section \ref{S:SMC} but we propose two
innovations which substantially improve the efficiency of the SMC
algorithm in Section \ref{S:ISPS}. The first innovation is to use
 coupled simulations of the Markov process allowing us to
consider a set of parameter values from the posterior distribution
using a single simulation. Coupled simulations for ABC were proposed
and successfully applied to household epidemic models in
\cite{Neal12} and we develop their usefulness for Markov processes
below. The second innovation is a simple, yet effective, importance
sampling procedure to {\it direct} the simulation process towards
the observed data. A similar idea, but very different in its
details, is the diffusion bridge used in \cite{GW11}. A key element
behind both innovations is to not significantly slow down the
simulation of the Markov process using the Gillespie algorithm which
is successfully achieved. Moreover, we show that the two innovations
substantially improve the efficiency of the SMC algorithm on their
own but that the real benefits are seen when they are combined. In
Section \ref{S:LV}, it is shown that using the coupled simulations
and importance sampling reduces the computational cost of the
algorithm by at least a factor of 30.

The paper is structured as follows. In Section \ref{S:MP}, we give
an introduction to Markov processes and a useful reparameterisation
of the Markov process which is exploited in developing the coupled
simulations. We also outline the two examples used to illustrate the
methodology, the Lotka-Volterra (predator-prey) model (\cite{BWK},
\cite{WKP}) and the  Repressilator model for gene regularity systems
(\cite{EL}, \cite{Toni09}). In Section \ref{S:SMC}, we outline the
SMC algorithm before introducing our two innovations coupled
simulations and importance sampling in Section \ref{S:ISPS}. In
Sections \ref{S:LV} and \ref{S:Rep}, we apply the SMC algorithm to
the Lotka-Volterra and Repressilator models, respectively. Parameter
estimation for the Lotka-Volterra model has proved to be
challenging, see \cite{BWK}, and it therefore gives a useful testing
ground for our methodology. The SMC algorithm works very effectively
even in the case where only the prey levels are observed. Analysing
the Repressilator model is substantially more challenging than the
Lotka-Volterra model due to the large number of events (between 1000
and 7000) which occur between observation points. However, the SMC
algorithm is successfully applied to this model. Finally, in Section
\ref{S:Conc} we give a brief summary of our findings and outline
possible extensions of the current work.

\section{Markov Processes and Examples} \label{S:MP}

In this Section we introduce Markov processes along with a simple
reparameterisation that will be exploited in developing coupled
simulations of the Markov process in Section \ref{S:ISPS}. We also
describe the two examples studied later in the paper the stochastic
Lotka-Volterra model (\cite{Wilkinson}, \cite{Toni09}, \cite{WKP})
and the Repressilator model (\cite{EL}, \cite{Toni09}). These
examples illustrate the chosen parameterisation framework.

Consider a Markov process $\mathcal{X}$ and let $\mathbf{X} (t)$
denote the state of the process at time $t$. Suppose that the
evolution of the Markov process is governed by the parameters
$\bthet_\alpha = (\balph, \bomega)$ and that there are $K$ possible
types of transitions. For $k=1,2,\ldots, K$, let $\varrho_k
(\mathbf{X} (t), \bthet_\alpha) = \alpha_k \rho_k (\mathbf{X} (t),
\bomega)$ denote the transition rate for a transition of type $k$,
given that the current state of the process is $\mathbf{X} (t)$ and
the parameters are $\bthet_\alpha$. For the generic description we
assume that the $\alpha_k$'s are distinct but this is not necessary
as demonstrated in the Repressilator example below. In many
situations $\bomega$ will be empty and then the parameters,
$\bthet_\alpha = \balph$, simply governs how fast events are taking
place. It will be useful later on to reparameterise the transition
rates by setting $\bthet = (\bbet, \bomega, \phi)$, where $\phi =
\alpha_K$ and for $k=1,2,\ldots, K$, $\beta_k = \alpha_k/\phi$.
(Note that $\beta_K =1$.) Then at time $t$, given that the current
state of the Markov process is $\mathbf{X} (t)$, the time until the
next event is ${\rm Exp} (\sum_{k=1}^K \varrho_k (\mathbf{X} (t),
\bthet_\alpha)) = {\rm Exp} (\sum_{k=1}^K \beta_k \rho_k (\mathbf{X}
(t), \bomega))/\phi$, where ${\rm Exp} (\lambda)$ denotes an
exponential distribution with mean $1/\lambda$. The probability that
the transition is of type $k$ is
\begin{eqnarray} \label{eq:rate1}
\frac{\varrho_k (\mathbf{X} (t), \bthet_\alpha)}{\sum_{l=1}^K
\varrho_l (\mathbf{X} (t), \bthet_\alpha)} = \frac{\beta_k \rho_k
(\mathbf{X} (t), \bomega)}{\sum_{l=1}^K \beta_l \rho_l (\mathbf{X}
(t), \bomega)}.
\end{eqnarray} The key observation is that ${\rm Exp} (\sum_{k=1}^K \beta_k \rho_k (\mathbf{X} (t),
\bomega))$ and \eqref{eq:rate1} are independent of $\phi$. Thus
$\phi$ denotes the speed measure of the Markov process; a fact that
we will exploit later in the paper.

The stochastic Lotka-Volterra model is a model for predator-prey
dynamics,  see, for example, \cite{Wilkinson}. There are two species
with for $t \geq 0$, $X_1 (t)$ and $X_2 (t)$ denoting the total
number of prey and predators at time $t$, respectively. There are
three types of transition; a birth of a prey ($(X_1(t), X_2(t))
\rightarrow (X_1(t)+1, X_2(t))$), a predator eats a prey, resulting
in the death of a prey and the birth of a predator  ($(X_1(t),
X_2(t)) \rightarrow (X_1(t)-1, X_2(t)+1)$) and the death of a
predator ($(X_1(t), X_2(t)) \rightarrow (X_1(t), X_2(t)-1)$). Let
$\bthet_\alpha = \balph = (\alpha_1, \alpha_2, \alpha_3)$ and
$\bthet = (\beta_1, \beta_2, \phi)$, where $\phi =\alpha_3$ and
$\beta_i = \alpha_i/\alpha_3$ $(i=1,2)$. Then  the infinitesimal
transition rates at time $t$ for the three types of transitions are
$\alpha_1 X_1 (t) = \beta_1 \phi X_1(t)$ (birth of a prey),
$\alpha_2 X_1(t) X_2(t) = \beta_2 \phi X_1(t) X_2 (t)$ (predator
eats prey) and $\alpha_3  X_2(t) = \phi X_2(t)$ (death of a
predator).


The Repressilator model is a popular toy model for gene regularity
systems (\cite{EL}, \cite{Toni09}). The model consists of three
genes which produce messenger RNA (mRNA), and where each gene's mRNA
transcribes a repressor protein  for the next gene in the loop.  For
$i=1,2,3$, let $Y_i (t)$ and $Z_i (t)$ denote the total abundance of
mRNA and protein, respectively, of gene $i$ at time $t$. The model
has 12 types of transitions. For each gene there is production
(birth) and decay (death) of mRNA and translation (birth) and decay
(death) of proteins. Let $\bthet_\alpha = (\alpha_1, \alpha_2,
\alpha_3, \alpha_4, \omega)$, or alternatively, $\bthet = (\beta_1,
\beta_2, \beta_3,\omega, \phi)$. Given $\mathbf{X} (t) = (Y_1 (t),
Y_2 (t), Y_3 (t), Z_1 (t), Z_2 (t), Z_3 (t))$, the transition rates
for gene $i$ ($i=1,2,3$) with correspondingly $j=3,1,2$, are,
\[ \begin{array}{lll}
\frac{\alpha_1}{1+ Z_j (t)^\omega} + \alpha_2 \left( = \phi \left\{
\frac{\beta_1}{1+ Z_j (t)^\omega}+ \beta_2 \right\} \right)  &
\mbox{ production of mRNA} & (Y_i (t) \rightarrow Y_i (t) +1), \\
\alpha_4 Y_i (t) (= \phi Y_i (t)) & \mbox{ decay of mRNA} &  (Y_i
(t) \rightarrow Y_i (t) -1), \\  \alpha_3 Y_i (t) (= \phi \beta_3
Y_i (t)) & \mbox{  protein translation} & (Z_i (t) \rightarrow Z_i
(t) +1)  \\ \alpha_3 Z_i (t)  (= \phi \beta_3 Z_i (t)) & \mbox{
protein decay } & (Z_i (t) \rightarrow Z_i (t) -1). \end{array} \]
Note that $\bomega$ is non-empty in this model with $\omega$
governing the effect of the repressor protein on the next gene in
the loop.

Throughout this paper we assume that the Markov process is observed
at discrete time points $t_0(=0), t_1, \ldots, t_n$. Typically, we
take the observation points to be equally spaced and at unit time
intervals so that $t_i =i$, although there is nothing to restrict us
to this case. Let $\mathbf{x}_i =\mathbf{X} (t_i)$ denote the state
of the system at time point $t_i$ and $\mathbf{x} = (\mathbf{x}_0,
\ldots, \mathbf{x}_n)$ with $\mathbf{x}_{a:b} =(\mathbf{x}_a,
\ldots, \mathbf{x}_b)$. We assume that the  process might be only
partially observed at any time point, that is, $\mathbf{X} (t_i) =
(\mathbf{Y} (t_i), \mathbf{Z} (t_i))$, where $\mathbf{Y} (t_i)$ is
observed and $\mathbf{Z} (t_i)$ is unobserved. For example, for the
analysis of the Repressilator model in Section \ref{S:Rep}, we
follow \cite{Toni09}, Section 3.2.1 in assuming that the  abundance
of mRNA is observed at each time point but that the protein levels
are unobserved. Therefore, we write $\mathbf{x}_i = (\mathbf{y}_i,
\mathbf{z}_i)$ to distinguish between the observed and unobserved
data at time point $i$, and we have that $\mathbf{y} =
(\mathbf{y}_0, \ldots, \mathbf{y}_n)$ denotes the observed data. We
are interested in $\pi (\bthet | \mathbf{y})$, the posterior
distribution of the parameters given the observed data. Note that
the approach we take allows, in principle, for different information
to be available at different time points. For example, in the
Lotka-Volterra model we could know both predator and prey numbers at
some time points and only prey numbers at other time points.
However, for ease of exposition, we shall restrict ourselves to
assuming that the same information about the Markov process is
observed at each time point with the possible exception of knowing
the full initial conditions of the Markov process.

\section{Sequential Monte Carlo} \label{S:SMC}

In this Section we outline a sequential Monte Carlo algorithm for
obtaining samples from $\pi (\bthet | \mathbf{y})$  or $\pi (\bthet,
\mathbf{z}_n |  \mathbf{y})$ with the latter being useful for
predictive purposes. The approach we take is based upon the Liu and
West filter,  \cite{LW}, which builds upon \cite{Gordon}.

Firstly, using Bayes' Theorem it is straightforward to show that,
for all $t=1,2,\ldots, n$,
\begin{eqnarray}  \label{eq:smclike1}
\pi (\bthet, \mathbf{z}_{0:t} | \mathbf{y}_{0:t}) \propto \pi
(\mathbf{y}_t, \mathbf{z}_t | \bthet, \mathbf{z}_{0:t-1},
\mathbf{y}_{0:t-1}) \pi ( \bthet, \mathbf{z}_{0:t-1}|
\mathbf{y}_{0:t-1}).
\end{eqnarray} For notational convenience, let $\pi (\bthet,
\mathbf{z}_{0:0} | \mathbf{y}_{0:0})$ denote the prior $\pi (\bthet,
\mathbf{z}_0)$ on $\bthet$ and $\mathbf{z}_0$. This is applicable,
for example when the initial condition $\mathbf{x}_0$ is set by
experimenter rather than $\mathbf{x}_0$ arising as the stationary
distribution of a Markov process. The marginal distribution of $\pi
(\theta, \mathbf{z}_t | \mathbf{y}_{0:t})$ can then be obtained by
integrating out $\mathbf{z}_{0:t-1}$. In the Markovian case this is
sufficient for studying the evolution of the process, since
\begin{eqnarray}  \label{eq:smclike2}
\pi (\bthet, \mathbf{z}_t, \mathbf{z}_{t-1} | \mathbf{y}_{0:t})
\propto \pi (\mathbf{y}_t, \mathbf{z}_t | \bthet, \mathbf{z}_{t-1},
\mathbf{y}_{t-1}) \pi ( \bthet, \mathbf{z}_{t-1}|
\mathbf{y}_{0:t-1}).
\end{eqnarray}
follows from  \eqref{eq:smclike1}  via simple algebraic
manipulation.

In principle \eqref{eq:smclike2} gives a sequential method for
computing the posterior distribution of $\{ \pi (\bthet|
\mathbf{y}_{0:t}) \}$. In particular, if we knew $\pi (\bthet,
\mathbf{z}_{t-1} | \mathbf{y}_{0:t-1})$ and $\pi (\mathbf{y}_t,
\mathbf{z}_t | \bthet, \mathbf{y}_{t-1}, \mathbf{z}_{t-1})$, we
could sample $(\bthet^1_{t-1}, \mathbf{z}_{t-1}^1), (\bthet^2_{t-1},
\mathbf{z}_{t-1}^2), \ldots, (\bthet^N_{t-1}, \mathbf{z}_{t-1}^N)$
from $\pi ( \bthet, \mathbf{z}_{t-1}| \mathbf{y}_{0:t-1})$ with
samples from the prior in the case $t=1$. Then we can sample
$\mathbf{z}_t^i$ from  $\pi ( \mathbf{z}_t | \bthet_{t-1}^i,
\mathbf{y}_{t-1}, \mathbf{z}_{t-1}^i)$ and compute $\varpi_{t-1}^i =
\pi (\mathbf{y}_t| \mathbf{z}_t^i, \bthet_{t-1}^i, \mathbf{y}_{t-1},
z_{t-1}^i)$, where, given the sampling scheme, $\varpi_{t-1}^i$ is
an unbiased probability weight for the trio $(\bthet_{t-1}^i,
\mathbf{z}_t^i, \mathbf{z}_{t-1}^i)$ being a sample from $\pi
(\bthet, \mathbf{z}_t, \mathbf{z}_{t-1} | \mathbf{y}_{0:t})$. The
process can be continued forward to $t+1$ by setting, for
$j=1,2,\ldots, N$, $(\bthet_t^j, \mathbf{z}_t^j) = (\bthet_{t-1}^i,
\mathbf{z}_t^i)$ with probability $\varpi_{t-1}^i/\sum_{k=1}^N
\varpi_{t-1}^k$.

It is well known that the above sequential importance resampling
(SIR) approach leads to degeneracy in the parameter space, $\Theta$,
see, for example, \cite{FT13}. Thus an alternative is needed for
generating $(\bthet_t^j, \mathbf{z}_t^j)$. A common approach,
\cite{Gordon}, is to perturb $\bthet_t^j$ by adding a small random
disturbance. For example, replacing $\bthet_t^j$ by $\bthet_t^j +
\bzet_t^j$, where $\bzet_t^j \sim N(\mathbf{0}, \mathbf{S}_t)$ for
some appropriately chosen variance matrix $\mathbf{S}_t$. As pointed
out in \cite{LW}, this approach leads to a loss of information
between time points, as artificial noise has been added to the model
parameters, which leads to increasingly diffuse estimates of the
posterior distribution. The solution proposed in \cite{LW}, Section
3.2, which we follow in this paper is as follows. Let $\bmu_{t-1}$
and $\mathbf{V}_{t-1}$ denote the estimated (Monte Carlo) posterior
mean and variance of $\pi (\bthet | \mathbf{y}_{0:t-1})$. Let $h >0$
be a smoothing parameter and $a = \sqrt{1-h^2}$. Simulate $\bzet_t^j
\sim N(\mathbf{0}, h^2\mathbf{V}_{t-1})$ and replace $\bthet_t^j$ by
\begin{eqnarray}  \label{eq:smclike3} a \bthet_t^j + (1-a) \bmu_{t-1} + \bzet_t^j,
\end{eqnarray} where $\bthet_t^j = \bthet_{t-1}^i$ with probability $\varpi_{t-1}^i/\sum_{k=1}^N
\varpi_{t-1}^k$. This ensures that the proposed parameters have mean
$\bmu_{t-1}$ and variance $\mathbf{V}_{t-1}$. It is recommended in
\cite{LW}, Section 3 to choose $\delta$ around 0.99, where $h^2= 1 -
((3 \delta -1)/2 \delta)^2$. This gives $h=0.1004$.

An alternative approach for updating $\bthet$ ($\bthet_\alpha$) is
to use MCMC moves for the particles, see for example, \cite{Storvik}
and \cite{Fearnhead}. Let $\mathbf{a}_{0:t}$ denote the trajectory
of the Markov process up to time $t$. For the case $\bomega =
\emptyset$ and with independent gamma distributed priors on the
components of $\balph$, $\pi (\bthet_\alpha | \mathbf{a}_{0:t},
\mathbf{x}_{0:t })$ is the product of $K$ independent Gamma
densities with
\begin{eqnarray} \label{eq:alpha} \alpha_k |  \mathbf{a}_{0:t}, \mathbf{x}_{0:t
} \sim {\rm Gamma} \left(E_k (t) +A_k, \int_0^t \rho_k (a_k (s)) \,
ds +B_k \right),
\end{eqnarray} where $E_k (t)$ denotes the total number of type $k$ transitions in the interval $[0,t]$
and ${\rm Gamma} (A_k, B_k)$ is the prior on $\alpha_k$. Therefore
it is sufficient to keep track of $(E_k (t) , \int_0^t \rho_k (a_k
(s)) \, ds)$ $(k=1,2,\ldots, K)$ rather than the full trajectory
$\mathbf{a}_{0:t}$, see \cite{Fearnhead}. The main reasons for
focussing on the \cite{LW} approach is the ease with which it can be
implemented with the coupled simulations introduced in Section
\ref{S:ISPS} using $\bthet$ rather than $\bthet_\alpha$ and that its
efficiency is not compromised by  $\bomega \neq \emptyset$, where no
low-dimensional sufficient statistics exist.

 In estimating the parameters
for discretely observed Markov processes, we start with a slight
adaption of the SIR algorithm with the Liu-West (\cite{LW}) filter
before developing improvements of the algorithm to make it more
efficient in Section \ref{S:ISPS}. First, note that $\pi
(\mathbf{y}_t, \mathbf{z}_t | \bthet, \mathbf{y}_{t-1},
\mathbf{z}_{t-1})$ is unknown. Therefore we simulate a realisation
of the Markov process with parameters $\bthet$ and starting  at
$\mathbf{X} (t-1) = (\mathbf{Y} (t-1), \mathbf{Z} (t-1)) =
(\mathbf{y}_{t-1}, \mathbf{z}_{t-1})$ between times $t-1$ and $t$.
If $\mathbf{Y} (t) = \mathbf{y}_t$, we set $\varpi =1$, the
simulation is consistent with the observed data, otherwise we set
$\varpi=0$. That is, we have an unbiased, indicator estimate for
$\pi (\mathbf{y}_t | \bthet, \mathbf{y}_{t-1}, \mathbf{z}_{t-1})$
with $\mathbf{z}_t$, in the case $\varpi_t =1$, being an unbiased
draw from $\pi ( \mathbf{z}_t | \mathbf{y}_t, \bthet,
\mathbf{y}_{t-1}, \mathbf{z}_{t-1})$. Secondly, if we run a fixed
number of particles $N$ it is possible that no simulation will be
accepted. Thus at each $t$ we sample particles until a fixed number
of simulations, $M$, are accepted. These ideas are developed further
in Section \ref{S:ISPS}, where the weights $\varpi$ are no longer
indicator variables and the simulations are run until a given
effective sample size, $M$, is reached.

The above developments give rise to the following sequential Monte
Carlo (SMC) algorithm.

{\bf SMC algorithm}
\begin{enumerate}
\item Set $l=0$.
\item While $l < M$:-
\begin{enumerate}
\item Sample $\bthet$ and  $\mathbf{z}_0$ from $\pi (\bthet, \mathbf{z}_0)$.
\item Simulate the Markov process $\mathcal{X}$ with parameters $\bthet$ and starting point $\mathbf{X} (t_0) = (\mathbf{y}_0, \mathbf{z}_0)$
from time $t_0$ to time $t_1$, and let $\mathbf{x}_1^\prime =
(\mathbf{y}_1^\prime, \mathbf{z}_1^\prime)$ denote $\mathbf{X}
(t_1)$.
\item If $\mathbf{y}_1^\prime = \mathbf{y}_1$, accept the simulation by setting $l=l+1$ and
$(\bthet_1^l,\mathbf{z}_1^l) = (\bthet,\mathbf{z}_1^\prime)$.
Otherwise the simulation is rejected.
\end{enumerate}
\end{enumerate}
Fix $h > 0$ and $a =\sqrt{1-h^2}$. For $t=2,3,\ldots,n$:-
\begin{enumerate}
\item Compute $\bmu_{i-1}$ and $\mathbf{V}_{t-1}$, the mean and variance of $\{ \bthet_{i-1}^1,\bthet_{i-1}^2, \ldots, \bthet_{i-1}^M\}$.
\item Set $l=0$.
\item While $l < M$:-
\begin{enumerate}
\item Sample $(\tilde{\bthet},\mathbf{z})$, uniformly at random from $\{ (\bthet_{i-1}^1,\mathbf{z}_{i-1}^1),
(\bthet_{i-1}^2,\mathbf{z}_{i-1}^2), \ldots,
(\bthet_{i-1}^M,\mathbf{z}_{i-1}^M)\}$.
\item Sample $\bzet \sim  N(\mathbf{0}, h^2\mathbf{V}_{t-1})$ and set $\bthet^\ast = a \tilde{\bthet} + (1-a) \bmu_{i-1} + \bzet$.
\item Simulate the Markov process $\mathcal{X}$ with parameters $\bthet^\ast$ and starting point $\mathbf{X} (t_{i-1})= (\mathbf{y}_{i-1}, \mathbf{z})$
from time $t_{i-1}$ to time $t_i$, and let $\mathbf{x}_i^\prime =
(\mathbf{y}_i^\prime, \mathbf{z}_i^\prime)$ denote $\mathbf{X}
(t_i)$.
\item If $\mathbf{y}_i^\prime = \mathbf{y}_i$, accept the simulation by setting $l=l+1$ and  $(\bthet_i^l,\mathbf{z}_i^l) = (\bthet^\ast,\mathbf{z}_i^\prime)$. Otherwise the simulation is rejected.
\end{enumerate}
\end{enumerate}

In the terminology of \cite{WKP}, this is a sequential exact
Bayesian computation (EBC) algorithm. The term EBC  refers to
employing an ABC simulation procedure without approximation,
{\i.e.}, an exact match is observed. It is straightforward  to adapt
the above algorithm to form a sequential ABC algorithm, where an
exact match is replaced by comparing summary statistics of the
simulated and observed data, accepting simulations where these are
sufficiently close, or to the case where the Markov process is
observed with observational error. In the sequel, we focus on
perfect, but partial, observation of the Markov process.

\section{Coupled simulations and Importance sampling} \label{S:ISPS}

The key limitation of the SMC algorithm described at the end of
Section \ref{S:SMC} is that the probability that a simulation, with
parameters $\bthet$ and starting from $\mathbf{X} (t_{i-1}) =
\mathbf{x}_{i-1}$, will result in $\mathbf{Y} (t_i) = \mathbf{y}_i$
is often prohibitively small. Therefore we look for ways to improve
the   rejection sampling scheme in  the SMC algorithm. Two
approaches are proposed, coupled simulation and importance sampling,
these can be used either in isolation or as in this paper combined.
However, for clarity of exposition we describe the two approaches
separately.

\subsection{Coupled simulation} \label{ss:coup}

For the coupled simulations we explicitly use the reparameterised
$\bthet = (\bbet, \bomega, \phi)$ for simulating the Markov process
$\mathcal{X}$. Let $\mathcal{W}$ denote the Markov process
$\mathcal{X}$ with parameters $\bvthet=(\bbet,\bomega,1)$, that is,
fixing $\phi=1$.  Thus $\mathcal{W}$ is a special case of
$\mathcal{X}$ but it suffices to study $\mathcal{W}$, since
$\mathcal{X}$ can be viewed as a speeded up $(\phi >1)$ or slowed
down $(\phi <1)$ version of $\mathcal{W}$. Let $\mathbf{W} (t) =
(\mathbf{W}^y (t), \mathbf{W}^z (t))$ denote the state of
$\mathcal{W}$ at time $t$, where $\mathbf{W}^y (t)$ and
$\mathbf{W}^z (t)$ correspond to the observed and unobserved
components, respectively, of the Markov process. Then at time $t$,
the transition rate for a transition of type $k$ is $\beta_k \rho_k
(\mathbf{W} (t), \bomega)$ $(k=1,2,\ldots, K)$. Now suppose that
$\mathbf{W} (0) = \mathbf{x}_{i-1}$. Then a realization of
$\mathbf{X} (t_i)$ given $\mathbf{X} (t_{i-1}) =\mathbf{x}_{i-1}$
and $\bthet = (\bbet, \bomega, \phi)$ can be obtained by setting
$\mathbf{X} (t_i) = \mathbf{W} ( \phi (t_i - t_{i-1}))$. Now if
$\mathcal{W}$ is simulated on the interval $[0,S]$ using
$\bvthet=(\bbet,\bomega,1)$, then realizations of $\mathbf{X} (t_i)$
can be generated for $\{ (\bbet, \omega, \phi); 0 \leq \phi \leq
S/(t_i -t_{i-1}) \}$. Therefore we can construct simulations from
$\mathbf{X} (t_i)$ for a whole set of parameters from a single
simulation of $\mathcal{W}$. In particular, for any $0 \leq \tau
\leq S$, where $\mathbf{W}^y (\tau) = \mathbf{y}_i$, we have a
simulated realisation with parameters $(\bbet, \bomega, \phi  =
\tau/(t_i - t_{i-1}))$ and starting at $\mathbf{X} (t_{i-1}) =
\mathbf{x}_{i-1}$ which results in $\mathbf{X} (t_i)$ with
$\mathbf{Y} (t_i) = \mathbf{y}_i$. Throughout this paper $t_i -
t_{i-1} =1$, so $\phi = \tau$ but this need not be the case.

We use the term coupled simulations for the above construction since
we are coupling together a sequence of parameter values in one
simulation. This is similar to the coupled ABC idea introduced in
\cite{Neal12}. The main difference is that here we  only consider
varying one parameter, $\phi$, rather than the whole set of
parameters, $\bthet$ in the coupling. Since $\mathbf{W} (t)$ is
piecewise constant, it is straightforward to obtain
$\mathcal{A}_\phi = \{ \phi; \mathbf{W}^y ( \phi (t_i - t_{i-1})) =
\mathbf{y}_i \}$. We discuss how to exploit $\mathcal{A}_\phi$ in
Section \ref{ss:smc} below.

\subsection{Importance sampling} \label{ss:imp}

A key problem with simulation-based  statistical inference is that
typically the probability that a simulated data set coincides with
the observed data set is extremely small. One natural solution is to
use importance sampling, \cite{Ripley}, to {\it steer} the
simulation towards the observed data, see \cite{NH13}. The aim is to
do this in a fast, efficient manner, so that the speed with which
the process is simulated is not severely compromised. The simple
solution we offer is to simulate from an alternative, time
inhomogeneous Markov process $\mathcal{R}$. Let $\mathbf{a} = \{
\mathbf{a}_s; t_{i-1} \leq s \leq t_i \}$ denote a realisation of a
Markov process between times $t_{i-1}$ and $t_i$ with
$f_{\mathcal{X}} (\mathbf{a}; \bthet, \mathbf{x}_{i-1})$ and
$f_{\mathcal{R}} (\mathbf{a}; \bthet, \mathbf{x}_{i-1})$ denoting
the probability density function for $\mathbf{a}$ under
$\mathcal{X}$ and $\mathcal{R}$, respectively, with parameters
$\bthet$ and starting at $\mathbf{x}_{i-1}$. We have that
\begin{eqnarray} \label{eq:is1}
&& \pi (\mathbf{Y} (t_i) = \mathbf{y}_i | \bthet, \mathbf{X} (t_{i-1}) = \mathbf{x}_{i-1}) \nonumber \\
&=& \int \pi (\mathbf{X} (t_i) = \mathbf{x}_i | \bthet, \mathbf{X}
(t_{i-1}) = \mathbf{x}_{i-1}, \mathcal{X} = \mathbf{a})
\frac{f_{\mathcal{X}} (\mathbf{a}; \bthet,
\mathbf{x}_{i-1})}{f_{\mathcal{R}} (\mathbf{a}; \bthet,
\mathbf{x}_{i-1})}
f_{\mathcal{R}} (\mathbf{a}; \bthet, \mathbf{x}_{i-1}) \, d \mathbf{a} \nonumber \\
&=& \int 1_{\{\mathbf{a}_{t_i}^y = \mathbf{y}_i \}}
\frac{f_{\mathcal{X}} (\mathbf{a}; \bthet,
\mathbf{x}_{i-1})}{f_{\mathcal{R}} (\mathbf{a}; \bthet,
\mathbf{x}_{i-1})} f_{\mathcal{R}} (\mathbf{a}; \bthet,
\mathbf{x}_{i-1}) \, d \mathbf{a}, \end{eqnarray} where
$\mathbf{a}_{t_i}^y$ corresponds to the observed components of the
Markov process. Thus using importance sampling we can simulate a
realisation $\mathbf{a}$ from $\mathcal{R}$, with
\begin{eqnarray} \label{eq:is2}
 1_{\{\mathbf{a}_{t_i}^y = \mathbf{y}_i \}}  \frac{f_{\mathcal{X}} (\mathbf{a}; \bthet, \mathbf{x}_{i-1})}{f_{\mathcal{R}} (\mathbf{a};
\bthet, \mathbf{x}_{i-1})}
\end{eqnarray} giving an unbiased estimate of $\pi (\mathbf{Y} (t_i) = \mathbf{y}_i | \bthet, \mathbf{X} (t_{i-1}) = \mathbf{x}_{i-1}) $.
Whilst, any choice of $\mathcal{R}$ could be used, for practical
purposes we want to be able to compute the ratio in \eqref{eq:is2}
 rapidly. Therefore we use the following Markov
process $\mathcal{R}$ with $\mathbf{R} (s)$ denoting the state of
the process at time $s$.
 Start with $\mathbf{R} (t_{i-1}) = \mathbf{x}_{i-1}$ and $P=1$. For
$s \geq t_{i-1}$, suppose that $\mathbf{R} (s) = \mathbf{r}$, then
the waiting time until the next event is exponentially distributed
with
rate $ \phi  \sum_{k=1}^K \beta_k \rho_k (\mathbf{r}, \bomega)$ 
and we simulate the time to the next event from this distribution.
Let $p_k =  \beta_k \rho_k (\mathbf{r}, \bomega) / \sum_{l=1}^K
\beta_l \rho_l (\mathbf{r}, \bomega)$, the probability that a type
$k$ transition takes place in $\mathcal{X}$, if $\mathbf{X} (s) =
\mathbf{r}$. Now instead of choosing the transition type according
to $\mathbf{p} =(p_1, \ldots, p_K)$, we choose according to
$\mathbf{q}= (q_1, \ldots, q_K)$, where $\mathbf{q}$ can depend upon
the current state $\mathbf{R} (s) = \mathbf{r}$, the time $s$, the
model parameters $\bthet$ and the target $\mathbf{y}_i$. If a
transition of type $k$ is chosen, we update $\mathbf{R} (s)$
accordingly and set $P = P \times p_k/q_k$. The process stops at
time $t$ with $P$ equal $f_{\mathcal{X}} (\mathbf{a}; \bthet,
\mathbf{x}_{i-1})/f_{\mathcal{R}} (\mathbf{a}; \bthet,
\mathbf{x}_{i-1})$. Thus the importance sampler is extremely easy to
implement and can offer significant gains in terms of efficiency of
the simulation algorithm. The choice of $\mathbf{q}$ is problem
specific and we discuss this in relation to the Lotka-Volterra and
Repressilator examples in Sections \ref{S:LV} and \ref{S:Rep},
respectively.

\subsection{Sequential Monte Carlo algorithm} \label{ss:smc}

We outline how the SMC algorithm introduced in Section \ref{S:SMC}
can be modified to make use of coupled simulations and importance
sampling. We begin by describing how to modify the sequential step
for $i >1$ before considering $i=1$.

Suppose that we have a sample of $N_{i-1}$ particles from $\pi (
\bthet, \mathbf{z}_{i-1}| \mathbf{y}_{0:i-1})$. That is, we have
$(\bthet_{i-1}^1, \mathbf{z}_{i-1}^1, \varpi_{i-1}^1), \ldots,$
$(\bthet_{i-1}^{N_{i-1}}, \mathbf{z}_{i-1}^{N_{i-1}},
\varpi_{i-1}^{N_{i-1}})$, where $\varpi_{i-1}^j$ is the relative
weight attached to $(\bthet_{i-1}^j, \mathbf{z}_{i-1}^j)$ and the
computation of $\varpi_{i-1}^j$ will be discussed below. As before
we can compute $\bmu_{i-1}$ and $\mathbf{V}_{i-1}$, although we will
primarily use $\tilde{\bmu}_{i-1}$ and $\tilde{\mathbf{V}}_{i-1}$,
the estimated mean and variance of $\bthet_{-\phi}$. For notational
convenience we denote $\bthet_{-\phi}$ by $\tilde{\bthet}$. It is
useful to write $\bthet = (\bbet, \bomega, \phi) (= (\tilde{\bthet},
\phi))$ with $\mathbf{V}_{i-1}$ written as
\begin{eqnarray} \label{eq:V1}
\mathbf{V}_{i-1} = \left( \begin{array}{ll} \tilde{\mathbf{V}}_{i-1}
& \tilde{C}_{i-1} \\
\tilde{C}_{i-1}^T & V_{i-1}^\phi
\end{array} \right).
\end{eqnarray}
Let $\sigma_\phi^2 = V_{i-1}^\phi - \tilde{C}_{i-1}^T
\tilde{\mathbf{V}}_{i-1}^{-1} \tilde{C}_{i-1}$, the conditional
variance of $\phi$ given the other parameters, $\tilde{\bthet}$.
This will be important in exploiting the coupled simulations.

Set $l=0$, $L=0$ and while $L < M$, we perform the following steps
in place of those in Step 3 of the SMC algorithm.
\begin{itemize}
\item[(a)] Sample $(\bthet,\mathbf{z})$ from $(\bthet_{i-1}^1, \mathbf{z}_{i-1}^1), \ldots,
(\bthet_{i-1}^{N_{i-1}}, \mathbf{z}_{i-1}^{N_{i-1}})$ with
probability $\varpi_{i-1}^j/ \sum_{k=1}^{N_{i-1}} \varpi_{i-1}^k$ of
choosing $(\bthet_{i-1}^j, \mathbf{z}_{i-1}^j)$.
\item[(b)] Sample $\tilde{\bzet} \sim N (0,h^2
\tilde{\mathbf{V}}_{i-1})$ and set $\bthet^\ast = a \tilde{\bthet} +
(1-a) \tilde{\mu}_{i-1} + \tilde{\bzet}$.

\item[(c)]
The conditional distribution of $\phi^\ast$ given $\bthet^\ast$ is
\begin{eqnarray} \label{eq:smc_1} N( a \phi + (1-a) \mu^\phi + \tilde{C}_{i-1}^T
\tilde{\mathbf{V}}^{-1}_{i-1} \tilde{\bzet}, \sigma_\phi^2) = N(
\tilde{\phi},  \sigma_\phi^2), \mbox{ say}. \end{eqnarray} Simulate
$U \sim U(0,1)$ and set $\mathcal{B}= \{ r; g (r;\tilde{\phi},
\sigma_\phi^2) \geq U g (\tilde{\phi}; \tilde{\phi}, \sigma_\phi^2)
\}$, where $g (r; \mu, \sigma^2)$ denotes the probability density
function of a $N(\mu, \sigma^2)$ evaluated at $r$.

\item[(d)] Set $P=1$ and $b_M = \sup \{ x; x \in \mathcal{B} \}$.

Simulate a Markov process $\mathcal{W}$ with parameters $\bvthet =
(\bthet^\ast,1)$ and $\mathbf{W} (0) = \mathbf{x}_{i-1}$ from time 0
to time $b_M$ incorporating importance sampling. That is, if
currently $\mathbf{W} (t) = \mathbf{w}$, we simulate $\tau$ from an
exponential distribution with rate
 $\sum_{k=1}^K \beta_k^\ast \rho_k (\mathbf{w},
\bomega^\ast)$. Then for $t \leq s < t + \tau$, $\mathbf{W} (s) =
\mathbf{w}$. At time $t + \tau$, a transition takes place with the
transition chosen according to $\mathbf{q}$. If a transition of type
$k$ is chosen, we update $\mathbf{W} (t+\tau)$ accordingly and set
$P = P \times p_k/q_k$, where $p_k = \beta_k^\ast \rho_k
(\mathbf{w}, \bomega^\ast) / \sum_{l=1}^K \beta_l^\ast \rho_l
(\mathbf{w}, \bomega^\ast)$.

For $s \in \mathcal{B}$, if $\mathbf{W}^y (s) = \mathbf{y}_i$, set
$k_s$ equal to the current value of $P$, otherwise set $k_s =0$.
Then set $\varpi = \int_{s \in \mathcal{B}} k_s \,ds$, the relative
weight of the simulation.

\item[(e)] If $\varpi>0$, sample $\phi^+$ from
$\mathcal{B}$ with probability density function proportional to
$k_s$. Set $l=l+1$ and $(\tilde{\bthet}_i^l, \phi_i^l,
\mathbf{z}_i^l, \varpi_i^l) = (\bthet^\ast, \phi^+, \mathbf{W}^z
(\phi^+), \varpi)$. Set $L= \{\sum_{k=1}^l \varpi_i^k \}^2/
\{\sum_{j=1}^l (\varpi_i^j)^2 \}$, the effective sample size.
\end{itemize}

We discuss the implications of the above procedure. Step (a) simply
draws a particle according to its relative weight from $\pi (\bthet,
\mathbf{z}_{i-1} | \mathbf{y}_{0:i-1})$ and step (b) applies the
Liu-West correction to all the parameters except $\phi$. Step (c)
produces a random set of values $\mathcal{B}$ of the form
$\mathcal{B} = \{ r \in [ \tilde{\phi} - T, \tilde{\phi} + T ]\}$,
where $T$ is a random variable determined by $U$ and
$\sigma^2_\phi$. For any $r \in \mathbb{R}$, the probability that $r
\in \mathcal{B}$ is proportional to $g (r; \tilde{\phi},
\sigma_\phi^2)$. The set $\mathcal{D} = \{ (\tilde{\bthet}, r,
\mathbf{z}); r \in \mathcal{B} \}$ is a set of parameters from the
distribution generated from $(\bthet_{i-1}^1, \mathbf{z}_{i-1}^1,
\varpi_{i-1}^1), \ldots, (\bthet_{i-1}^{N_{i-1}},
\mathbf{z}_{i-1}^{N_{i-1}},\varpi_{i-1}^{N_{i-1}})$ with the
Liu-West Gaussian kernel smoothing. This mimics the posterior sets
generated in \cite{Neal12} with more details on the construction of
sets for a random variable given in the Appendix. In particular,
$\mathcal{D}$ is an (approximate) sample of parameter values from
$\pi ( \bthet, \mathbf{z}_{i-1}| \mathbf{y}_{0:i-1})$ with the
approximation given by the Liu-West smoothing and is no different to
that generated by the SMC algorithm. Returning to
\eqref{eq:smclike2}, we have a sample from $\pi ( \tilde{\bthet},
\mathbf{z}_{i-1}| \mathbf{y}_{0:i-1})$ and have constructed a set
$\mathcal{B}$ from $\pi (\phi | \tilde{\bthet}, \mathbf{z}_{i-1},
\mathbf{y}_{0:i-1})$. Therefore we need to estimate $\pi
(\mathbf{y}_i, \mathbf{z}_i | \tilde{\bthet}, \mathbf{x}_{i-1})$ in
order to get a sample from $\pi (\bthet, \mathbf{z}_i
|\mathbf{y}_{0:i})$. In step (d), we simulate the process
$\mathcal{W}$ and consider $\phi$ values lying in $\mathcal{B}$. We
simultaneously consider realisations of $\mathcal{X}$ for all
parameters $\{(\bthet^\ast,\phi); \phi \in \mathcal{B} \}$ and it
thus suffices to simulate $\mathcal{W}$ on the interval $[0, b_M]$.
Note that $\varpi$ is given by
\begin{eqnarray} \label{eq:pi:2} \int_{\phi \in \mathcal{B}} 1_{ \{\mathbf{W}_\phi^y = \mathbf{y}_t  \}} \frac{f_{\mathcal{W}} (\mathbf{a}; (\tilde{\bthet}^\ast, \phi),
\mathbf{x}_{t-1})}{f_{\mathcal{R}} (\mathbf{a};
(\tilde{\bthet}^\ast, \phi), \mathbf{x}_{t-1})} \, d\phi =
\int_{\phi \in \mathcal{B}} k_\phi \, d\phi. \end{eqnarray} Now
$\varpi/(\sqrt{2 \pi} \sigma_\phi)$ is an unbiased estimate of $\pi
(\mathbf{Y} (t_i) = \mathbf{y}_i | \tilde{\bthet}^\ast, \mathbf{X}
(t_{i-1}) = \mathbf{x}_{i-1})$, where
\begin{eqnarray} \label{eq:pi:1}
&& \pi (\mathbf{Y} (t_i) = \mathbf{y}_i | \tilde{\bthet}^\ast,
\mathbf{X} (t_{i-1}) = \mathbf{x}_{i-1}) \nonumber \\ &=& \int \int
\pi (\mathbf{Y} (t_i) = \mathbf{y}_i | \tilde{\bthet}^\ast,
\mathbf{X} (t_{i-1}) = \mathbf{x}_{i-1}, \phi, \mathcal{W} =
\mathbf{a}) \frac{f_{\mathcal{W}} (\mathbf{a}; (\tilde{\bthet}^\ast,
\phi), \mathbf{x}_{i-1})}{f_{\mathcal{R}} (\mathbf{a};
(\tilde{\bthet}^\ast, \phi), \mathbf{x}_{i-1})} \nonumber \\ &&
\times f_{\mathcal{R}} (\mathbf{a}; (\tilde{\bthet}^\ast, \phi),
\mathbf{x}_{i-1}) \pi (\phi | \tilde{\bthet}^\ast, \mathbf{X}
(t_{i-1})=
\mathbf{x}_{i-1}) \, d\mathbf{a} \, d\phi \nonumber \\
&=& \int \int k_\phi f_{\mathcal{R}} (\mathbf{a};
(\tilde{\bthet}^\ast, \phi), \mathbf{x}_{i-1}) \pi (\phi |
\tilde{\bthet}^\ast, \mathbf{X} (t_{i-1})= \mathbf{x}_{i-1}) \,
d\mathbf{a} \, d\phi.
\end{eqnarray}
The details are given in the Appendix. The computation and storage
of $k_s$ $(s \in \mathcal{B})$ is straightforward as $k_s$ is
piecewise-constant. Then $\varpi/(\sqrt{2 \pi} \sigma_\phi )$ is an
estimate of $\pi (\tilde{\bthet}, \mathbf{z}_{i-1} |
\mathbf{y}_{0:i})$, with the computation of $\varpi$ based on the
simulation taking into account both the importance sampling weights
(steering of the simulation) and the time spent $\mathbf{W}^y (s) =
\mathbf{y}_i$ for $s \in \mathcal{B}$. Finally, in step (e), we
obtain a sample $(\phi, \mathbf{z}_i)$ from $\pi (\phi, \mathbf{z}_i
| \tilde{\bthet}, z_{i-1}, \mathbf{y}_{0:i})$. This is done on the
basis of the simulated $\mathcal{W}$ by sampling $\phi^+$ from the
set $\mathcal{B}$, proportional to $k_s$, and then setting
$\mathbf{z}_i = \mathbf{W}^z (\phi^+)$, the corresponding value of
the process for the unobserved components of the Markov process.

For the case $i=1$, the choice of $(\tilde{\bthet}, z)$ and
$\mathcal{B}$ changes to take into account the prior distribution
but steps (d) and (e) remain unchanged. If we have that $\pi
(\bthet, \mathbf{z}) = \pi (\tilde{\bthet}, \mathbf{z}) \pi (\phi)$,
then we simply simulate $(\tilde{\bthet}, \mathbf{z})$ from its
prior and set $\mathcal{B} = \{ r; \pi (\phi =r) \geq U \max_x \pi
(\phi =x) \}$, where $U \sim U(0,1)$, and proceed as above. However,
a prior may naturally be specified in terms of $\bthet_\alpha$ and
the above prior independence between $\tilde{\bthet}$ and $\phi$
will then not be the case in general. In this paper we consider the
case where the prior on $\bthet_\alpha$ is uniform on $D_\alpha
\subset \mathbb{R}^d$ with $D_\alpha$ being a $d$-dimensional cube.
This results in the prior on $\bthet$ being uniform on a set  $D
\subset \mathbb{R}^d$ and it is then easy to simulate
$\tilde{\bthet}$ and choose the appropriate $\mathcal{B}$. We
discuss the details in relation to specific examples in Sections
\ref{S:LV} and \ref{S:Rep}.

A key question is how much more computationally intensive is the
sequential Monte Carlo algorithm with coupled simulations and
importance sampling compared with the SMC algorithm. The
computationally intensive part of both algorithms is running the
simulations with the computations of means, variances and other
quantities between time points being minimal. Therefore we compare
mean time required per simulation. For $i >1$, the mean period
length for which the SMC algorithm is run is $\tilde{\phi}$ and for
the sequential Monte Carlo algorithm with coupled simulations the
mean period length is $\tilde{\phi} + 1.254 h \sigma_\phi$.
Typically, $\sigma_\phi$ is relatively small compared with
$\tilde{\phi}$, so the additional time required per simulation is
small. Furthermore, if $\sigma_\phi$ is relatively large, then so
typically will $\mathcal{B}$, and the use of coupled simulations
will be particularly useful. The computation of importance sampling
probabilities depends upon how these are computed but for the
examples in this paper, the computation of $\mathbf{q}$ is similar
in complexity to the computation of $\mathbf{p}$. Therefore
incorporating coupled simulations and importance sampling will at
most double the time required per simulation. For the examples
studied in Sections \ref{S:LV} and \ref{S:Rep} it was found that the
additional time was approximately only $20\%$ longer per simulation.

A secondary question is the choice of $h$. For \cite{LW}, the choice
of $h$ depends upon kernel smoothing considerations, a compromise
between under and over smoothing with the ideal $h \approx 0.1$. For
the  sequential Monte Carlo algorithm, additionally $h$ determines
the size of the set $\mathcal{B}$ in the coupled simulations, and
increasing $h$ will increase the acceptance rate. Thus alongside
increasing $h$, we can increase $M$, the effective sample size
without increasing the mean number of simulations at each time
point. Consequently, we typically take $h$ in the range $0.15$ to
$0.20$, which still results in $\delta$ between $0.95-0.99$, the
range advocated in \cite{LW}. We briefly discuss varying $h$ at the
end of Section \ref{S:LV:full}.

\section{Lotka-Volterra model} \label{S:LV}

\subsection{Introduction} \label{S:LV:intro}

The stochastic Lotka-Volterra model has proved a useful testing
ground for statistical inference techniques for Markov processes.
For example, reversible jump MCMC (\cite{BWK}), SMC-ABC
(\cite{Toni09}), particle MCMC (\cite{GW11}) and piecewise ABC
(\cite{WKP}). In all of the above papers the methodology is tested
on simulated data and it is assumed that the Lotka-Volterra process
is observed at a discrete number of points with either both predator
and prey numbers being observed or only prey numbers are observed.
The observations are assumed to be exact in \cite{BWK} and
\cite{WKP}, to have observation error in \cite{GW11} and are
averaged over replicates in \cite{Toni09}. The reversible jump MCMC
algorithm of \cite{BWK} is computationally intensive and experiences
poor mixing due to the large amount of data augmentation involved.
The SMC-ABC algorithm of \cite{Toni09} appears to work reasonably
with multiple data replicates with the true parameter values lying
in the support of the reported posterior distribution. However, it
is not possible to assess the level of approximation of the
posterior distribution. The particle MCMC of \cite{GW11}, which uses
an SDE approximation and diffusion bridges (importance sampling),
works well when the data is observed with error. However, the
performance of the particle MCMC severely worsens as the noise term
becomes smaller, see \cite{WKP}. The piecewise ABC of \cite{WKP}
requires that both predator and prey numbers are observed and its
performance is highly sensitive to the choice of prior.

We consider the case where the observations are assumed to be exact
with either both predator and prey numbers or only prey numbers
observed. It is relatively straightforward to adapt the methods to
the case with observational error. We present analysis from one
simulated data set although similar findings were observed with
other data sets across a range of parameter values. The data
consists of observations at time points $t=0,1, \ldots, 40$, of a
simulation of the Lotka-Volterra model with with $\bthet_\alpha =
(1,0.005,0.6)$ $(\bthet = (\beta_1, \beta_2, \phi)=
(5/3,5/600,3/5))$ and $\mathbf{x}_0 = (71,79)$. The observed data
are plotted in Figure 1. The parameter values chosen correspond to
those used in \cite{WKP} and are double the parameters values used
in \cite{BWK}. Thus our observations are further apart than
\cite{BWK}, with \cite{WKP} not reporting the observation times. The
total number of events in the simulation between time 0 and time 40
is over 14000, which highlights the degree of data augmentation that
would be required if a data-augmentation MCMC algorithm were to be
used. A vague prior was placed on $\bthet$ with $U(-4,2)$ and
$U(-8,-3)$ chosen for $\log(\beta_1)$ and $\log(\beta_2)$,
respectively, and a $U(0,2)$ prior for $\phi$.

\begin{figure}
\epsfig{file=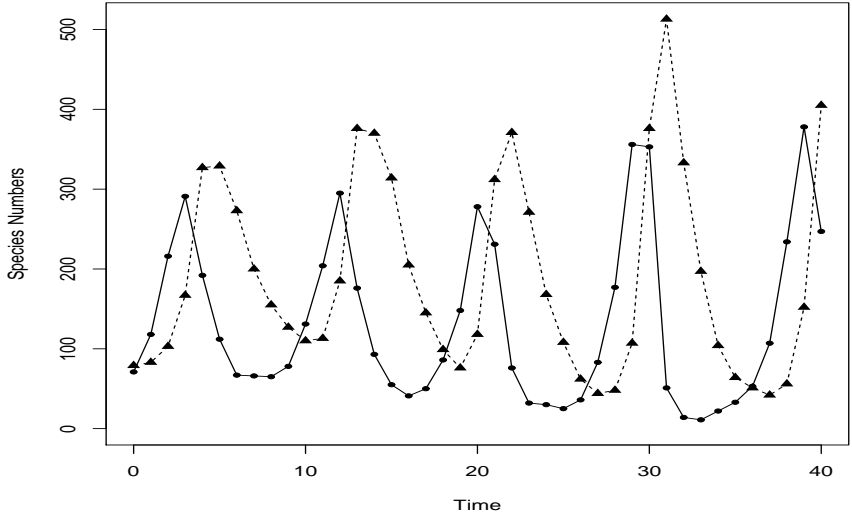, height=8cm, width=12cm}
\caption{Circles (solid line) and triangles (dashed line) denote
total numbers of prey and predators, respectively, at the
observation points.} \label{fig:1}
\end{figure}


\subsection{Predator and prey numbers observed} \label{S:LV:full}

The implementation of the SMC algorithm is as detailed in Section
\ref{ss:smc} with only details of the importance sampling, given
below, needing to be specified. In particular, we use a local
linearisation of the Markov process to devise the importance
sampling distribution $\mathbf{q}$.

For interval $i$, the {\it target} (observed) data is
$\mathbf{x}_i$. In particular, given parameters $\bvthet^\ast$ and
$\phi | \bvthet^\ast \sim N(\tilde{\phi}, \sigma_\phi^2)$, we aim
for $\mathbf{W} (\tilde{\phi}) = \mathbf{x}_i$, where $\mathbf{W}
(0) = \mathbf{x}_{i-1}$. Given an event occurs at time $0 < s <
\tilde{\phi}$ and $\mathbf{W} (s) = \mathbf{w}$, we choose
$\mathbf{q}$ as follows, with $\mathbf{q} = \mathbf{p}$ for $s \geq
\tilde{\phi}$. Let $L_1 = x_{i,1} - W_1 (s)$ and $L_2 = x_{i,2} -
W_2 (s)$, the differences between the {\it target} and the current
prey and predator numbers, respectively. Let $R = (\tilde{\phi} -s)
\times \sum_{k=1}^K \beta_k \rho_k(\mathbf{W} (s))$, the expected
number of events in $\mathcal{W}$ on the interval
$(s,\tilde{\phi}]$, if the current transition rates are maintained.
Then we set
\begin{eqnarray} \label{eq:lv_1}
Q_1 &=& \max \left\{ 0, \frac{R + 2L_1 + L_2}{3R} \right\}  \nonumber \\
Q_2 &=& \max \left\{ 0,\frac{R - L_1 + L_2}{3R} \right\}  \nonumber \\
Q_3 &=& \max \left\{ 0,\frac{R - L_1 - 2 L_2}{3R} \right\},
\end{eqnarray}
and then normalise, if necessary, by setting $Q_i$ equal to
$Q_i/\sum_{j=1}^3 Q_j$. Assuming that $Q_1, Q_2$ and $Q_3$ in
\eqref{eq:lv_1} are positive then the average effect over the
interval $(s,\tilde{\phi}]$ with $R$ transitions is for the number
of prey and predator to increase by $L_1$ and $L_2$ (decrease if
$L_1/L_2$ are negative), respectively. Thus naively we could set
$q_i = Q_i$ $(i=1,2,3)$. However, this leads to a very poorly
performing importance sampler. We found it best to put more weight
on $\mathbf{Q} = (Q_1, Q_2, Q_3)$ as $s$ approached $\tilde{\phi}$
with
\begin{eqnarray} \label{eq:lv_2}
q_i = \left(1- \epsilon \left(\frac{s}{\tilde{\phi}} \right)^\kappa
\right) p_i + \epsilon \left(\frac{s}{\tilde{\phi}} \right)^\kappa
Q_i \hspace{0.5cm} (i=1,2,3),
\end{eqnarray}
for some $0 \leq \epsilon \leq 1$ and $\kappa \geq 0$. We found that
$\epsilon =0.3$ and $\kappa =2$ performed well across a range of
data sets as a compromise between the transition probability,
$\mathbf{p}$, and the steering probability $\mathbf{Q}$.

We set $M=1000$ and $h=0.15$. We ran the code with no importance
sampling for $i=1$ and the above importance sampling regime with
$\epsilon =0.3$ and $\kappa =2$ for $i >1$. It was observed that it
was beneficial not to have importance sampling at the first time
point. The total number of simulations across the 40 time points was
14,100,811, a mean of just over 350,000 simulations per time point.
There is considerable variation in the number of simulations per
time point ranging from 54,447 for time point 8 to 2,516,218 for
time point 4. In Figure 2, the estimated posterior mean plus and
minus two times the estimated posterior standard deviation of
$\alpha_i (= \beta_i \phi)$ $(i=1,2,3)$, evaluated after each time
point. We note a significant change at time point 4 and also notable
changes at time points 21 and 31 which are the other two time points
that required over a million simulations. However, it is not obvious
from the data in Figure 1 that we should expect a notable change in
the parameters at these time points. Finally, the estimated
posterior means and standard deviations for $\bthet_\alpha |
\mathbf{x}_{0:40}$ are given in Table \ref{table:1a}. The posterior
means are close to the chosen parameter values. The standard
deviations are similar to those reported in \cite{BWK}, Table 1,
using reversible jump MCMC, admittedly for a different data set, and
this is observed across different data sets. Thus the Liu-West
procedure is not only providing good estimates of the mean of the
parameters but also the uncertainty in the posterior distribution of
the parameters.

\begin{figure}
\epsfig{file=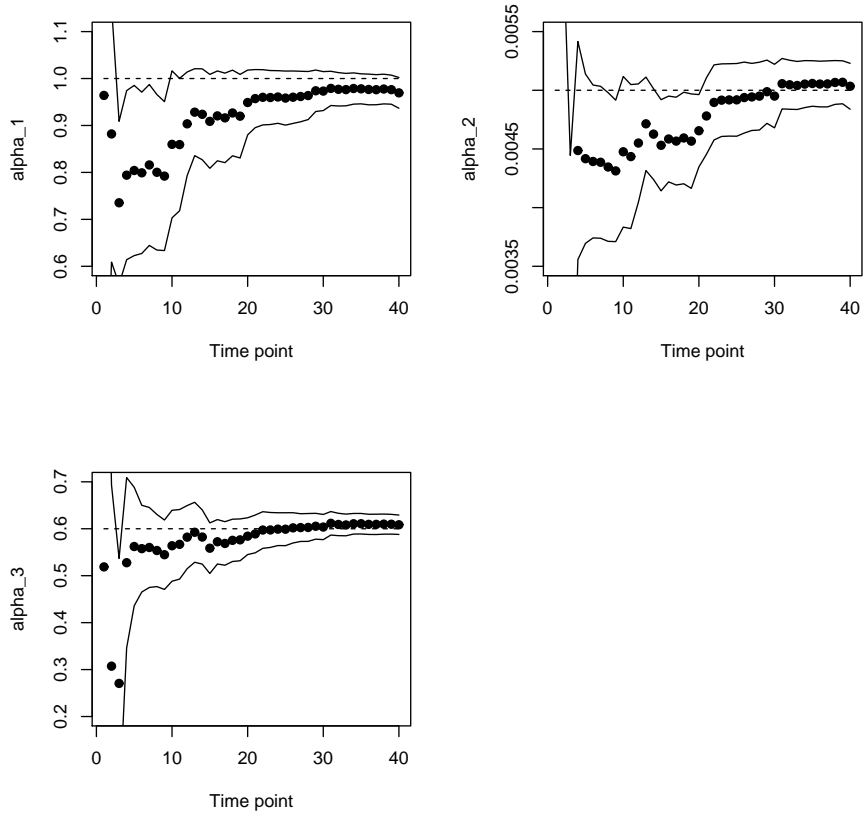, height=12  cm, width=12cm}
\psfrag{alpha_1}[][]{$\alpha_1$} \psfrag{alpha_2}[][]{$\alpha_2$}
\psfrag{alpha_3}[][]{$\alpha_3$}
\caption{Plots of mean of $\theta_\alpha$ parameters plus/minus two
standard deviations.} \label{fig:2}
\end{figure}

\begin{table}
{Table 1: Estimated posterior means and standard deviations for
$\bthet_\alpha | \mathbf{x}_{0:40}$.}
{%
\begin{tabular}{|c|c|c|c|}
\hline Parameter & $\alpha_1$ &
$\alpha_2$ & $\alpha_3$ \\
\hline  Mean &  0.970 & 0.00503 & 0.609 \\
\hline St. Dev. & $1.64 \times 10^{-2}$ & $9.78 \times 10^{-5}$ &
$1.03 \times 10^{-2}$ \\ \hline
\end{tabular}}
\label{table:1a}
\end{table}

It is informative to compare the performance of the SMC algorithm
with  coupled simulations and importance sampling with the SMC
algorithm with only one or neither of these modifications. The
coupled simulations required on average twice as long to run at the
first time point. However for subsequent time points the additional
computational cost was substantially smaller following the
discussion at the end of Section \ref{S:ISPS}. The importance
sampling slowed down the speed of the simulations by at most $20\%$.
It was found that for $M=1000$, the SMC algorithm with neither
modification required approximately 920 million simulations in
total, whereas the SMC algorithm with only coupled simulations or
importance sampling required approximately 90 million simulations in
total in both cases. Thus the modifications to the SMC algorithm
make it at least 30 times faster  (allowing for twice as long per
simulation).

Finally, we comment briefly on varying $h$. We found that increasing
$h$ to $0.20$ or reducing $h$ to $0.10$, for fixed $M$ resulted in
approximately $28\%$ fewer and $22 \%$ more simulations,
respectively. Consistent estimation of the posterior means was
observed across the different values of $h$ with the estimated
posterior standard deviation increasing slightly with increasing
$h$.

\subsection{Only prey levels observed} \label{S:LV:prey}

A more challenging statistical problem is where only the prey
numbers are observed at each time point, \cite{BWK}. The reversible
jump MCMC algorithm of \cite{BWK} incurs additional mixing problems
with this case but is still able to recover parameter values
consistent with those used for simulation, see \cite{BWK}, Table 2.
The piecewise ABC algorithm of \cite{WKP} is unable to deal with
this case as complete observation of the Markov process at each time
point is required.

Implementation of the SMC algorithm is similar to in Section
\ref{S:LV:full} with a few minor modifications. The same prior is
used for $\bthet$ but now a prior is required for $X_2(0)$, for
which we use a discrete uniform on the range 10 to 300, inclusive.
Given that we only require the simulations to match on prey levels,
we increased $M$ to 10000 and reduced $h$ to 0.1. This resulted in a
total of 8,991,017 simulations over the 40 time points. Finally, the
importance sampling is modified to take into account that only the
prey numbers need to match. Specifically, for $0 < s <
\tilde{\phi}$, we let $L_1 = x_{i,1} - W_1 (s)$ and $R=
(\tilde{\phi} -s) \times \sum_{k=1}^K \beta_k \rho_k(\mathbf{W}
(s))$ as before. Then we set,
\begin{eqnarray} \label{eq:lv_3}
Q_1 &=&  \frac{R (p_1 +p_2) +L_1}{2R}  \nonumber \\
Q_2 &=& \frac{R (p_1 +p_2) -L_1}{2R},
\end{eqnarray}
restricted to $0 \leq Q_1, Q_2 \leq p_1 + p_2$. That is, if $Q_1 <0$
$(Q_2 <0)$, we set $Q_1 = 0$ $(Q_1 = p_1 + p_2)$ and $Q_2 = p_1 +
p_2$ $(Q_2=0)$. Thus letting $Q_3 = p_3$, we set
\begin{eqnarray} \label{eq:lv_4}
q_i = \left(1- \epsilon \left(\frac{s}{\tilde{\phi}} \right)^\kappa
\right) p_i + \epsilon \left(\frac{s}{\tilde{\phi}} \right)^\kappa
Q_i \hspace{0.5cm} (i=1,2,3),
\end{eqnarray}
with $\epsilon =0.3$ and $\kappa =2$ as before. Note that $q_3 =
p_3$.

In Figure 3, the estimated posterior mean plus and minus two times
the estimated posterior standard deviation of $\alpha_i (= \beta_i
\phi)$ $(i=1,2,3)$, evaluated after each time point. We again note a
significant change in the parameters at time points 21 and 31. The
estimation of $\alpha_1$ is more erratic than the other two
parameters but appears to be settling down towards the end of the
observation period. The estimated posterior means and standard
deviations for $\bthet_\alpha | \mathbf{y}_{0:40}$ are given in
Table \ref{table:1b}. Whilst the estimated posterior means are
similar to those obtained in Table \ref{table:1a} with predator and
prey numbers observed, there is substantially greater uncertainty in
the posterior distribution of the parameters. This is consistent
with \cite{BWK}, Table 2.

\begin{figure}
\epsfig{file=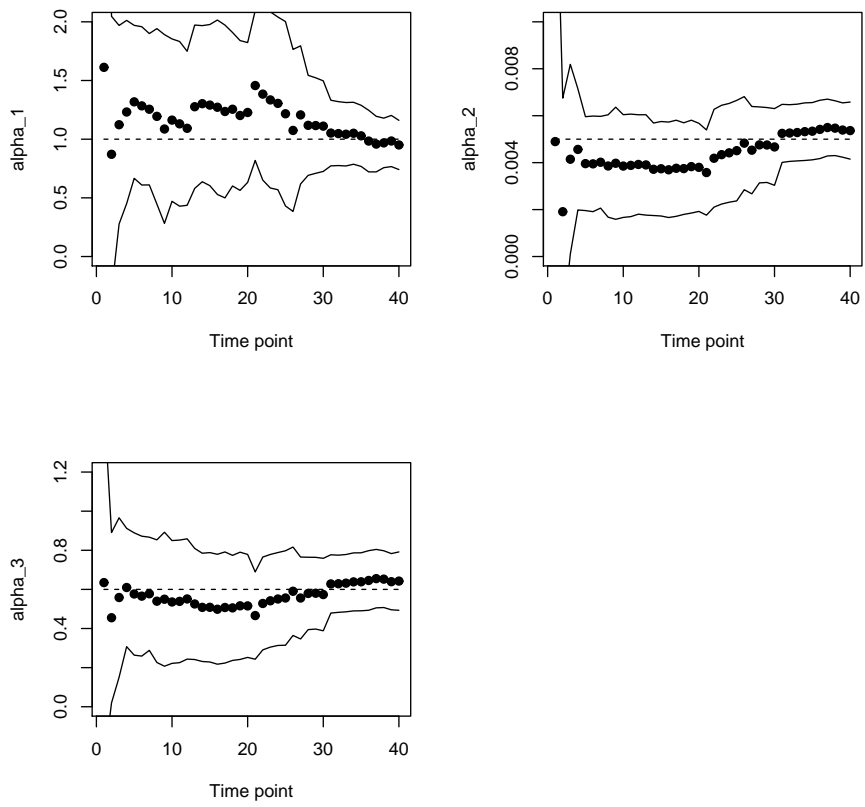, height=12cm, width=12cm}
\psfrag{alpha_1}[][]{$\alpha_1$} \psfrag{alpha_2}[][]{$\alpha_2$}
\psfrag{alpha_3}[][]{$\alpha_3$}
\caption{Plots of mean of $\theta_\alpha$ parameters plus/minus two
standard deviations.} \label{fig:3}
\end{figure}

\begin{table}
{Table 2: Estimated posterior means and standard deviations for
$\bthet_\alpha | \mathbf{y}_{0:40}$.}
{%
\begin{tabular}{|c|c|c|c|}
\hline Parameter & $\alpha_1$ &
$\alpha_2$ & $\alpha_3$ \\
\hline  Mean &  0.951 & 0.00537 & 0.642 \\
\hline St. Dev. & $1.05 \times 10^{-1}$ & $6.08 \times 10^{-4}$ &
$7.46 \times 10^{-2}$ \\ \hline
\end{tabular}}
\label{table:1b}
\end{table}

\section{Repressilator model} \label{S:Rep}

We follow \cite{Toni09} in analysing data simulated from the
Repressilator model with $\bthet_\alpha = (\alpha_1, \alpha_2,
\alpha_3, \alpha_4, \omega) = (1000,1,5,1,2)$, initial mRNA levels
$\mathbf{Y} (0) = (0,0,0)$ and protein levels $\mathbf{Z} (0) =
(2,1,3)$. The data was simulated for 50 time units with over 140,000
events taking place. The mRNA levels of the three genes were
observed at times $1,2, \ldots, 50$, shown in Figure \ref{fig:4}
below, whilst the protein levels, apart from the initial numbers,
were unobserved.

\begin{figure}
\epsfig{file=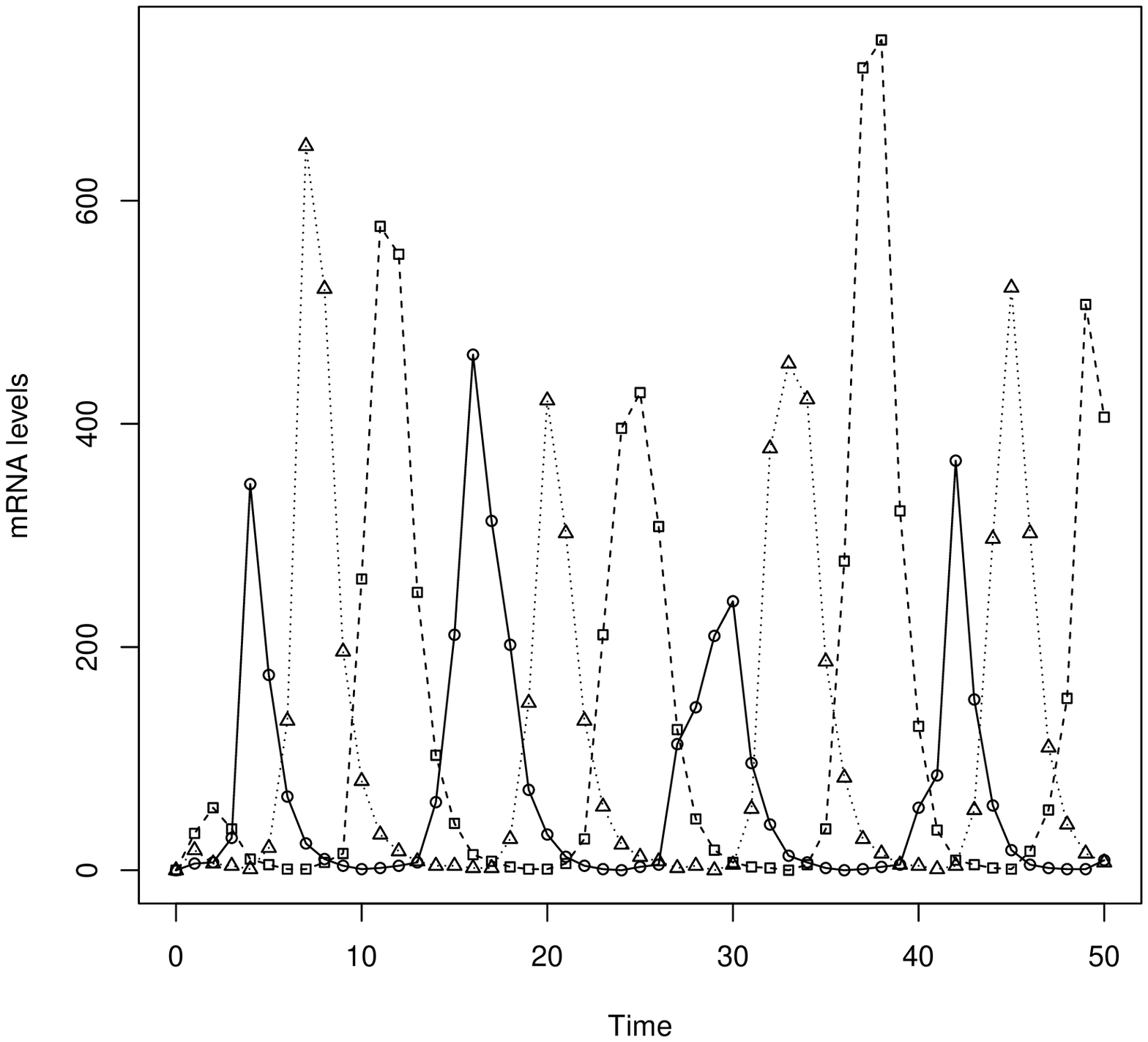, height=8cm, width=12cm}
\caption{Plot of observations of mRNA levels; gene 1 (circles -
solid line), gene 2 (squares - dashed line) and gene 3 (triangles -
dotted line).} \label{fig:4}
\end{figure}

The key difference from \cite{Toni09} is that we assume that
$\alpha_4$ is unknown. Uniform priors are chosen for
$\bthet_\alpha$: $\pi (\alpha_1) \sim U(500,2500)$, $\pi (\alpha_2)
\sim U(0,10)$, $\pi (\alpha_3) \sim U(0,10)$, $\pi (\alpha_4) \sim
U(0.5,2)$ and $\pi (\omega) \sim U(0,10)$. Transforming this into a
prior for $\bthet$ is straightforward, by drawing $\beta_1 \sim
U[250,5000]$, $\beta_2 \sim U[0,10]$, $\beta_3 \sim U[0,20]$ and
$\omega \sim U[0,10]$. Then set $\mathcal{B} \subseteq [0.5,2]$ such
that for $\phi \in \mathcal{B}$, $\alpha_k = \beta_k \phi$
$(k=1,2,3)$ is within the appropriate prior range.

Employing the SMC algorithm for the Repressilator model is more
computationally challenging than for the Lotka-Volterra data set.
Firstly, the simulated data has on average over 2800 events between
observations. Secondly, the observed data is 3 dimensional rather
than 1 or 2 dimensional as in the Lotka-Volterra case. Thirdly,
there are 12 rather than 3 transition types. Consequently, we found
taking $M=1000$ and $h=0.2$ $(a=0.9798)$ offered a good compromise
between precision of estimates and efficient running of the SMC
algorithm.

In Section \ref{S:LV} for the Lotka-Volterra model, a locally linear
importance sampling scheme was found to be useful. Given the
non-linear behaviour of the growth and decline of the mRNA gene
levels, an alternative approach is used here. Suppose that the
target for time $t_i$ is $\mathbf{y}_i = (y_{i,1}, y_{i,2},
y_{i,3})$ with correspondingly $\mathbf{z}_i = (z_{i,1}, z_{i,2},
z_{i,3})$ unobserved and that the process is currently at
$\mathbf{W} (s) = (\mathbf{w}_s^y, \mathbf{w}_s^z)$ with a
transition occurring at time $s$. Let $p^P_k$, $p^D_k$, $p^T_k$ and
$p^C_k$ denote the transition probabilities of mRNA production, mRNA
decay, protein translation and protein decay, respectively, of gene
$k$. Set $q^T_k = p^T_k$ and $q^C_k = p^C_k$. That is, we focus the
importance sampling on mRNA production and decay where we have a
target leaving the protein probabilities unchanged. Let
\begin{eqnarray} \label{eq:rep_1}
D_k = \left\{\frac{\beta_1}{1 + y_{i,j}^\omega} + \beta_2 + y_{i,k}
\right\}^{\frac{1}{2}} \times \left\{\frac{\beta_1}{1 +
(w_{s,j}^z)^\omega} + \beta_2 + w_{s,k}^y \right\}^{\frac{1}{2}},
\end{eqnarray} where $j=3,1,2$ corresponds to $k=1,2,3$. Then $D_k$
is geometric mean of the rate of change (production and decay) of
gene $k$ mRNA at times $s$ (current) and $t$ (target). Note that
since $z_{i,j}$ is unobserved $y_{i,j}$ represents a {\it best
guess} for $z_{i,j}$. Let $L_k = y_{i,k} - w_{s,k}^y$, the
difference between the target and current levels of gene $k$ mRNA.
Let $\delta = \tilde{\phi} -s$, then we set $b_k = (\delta D_k +
L_k)/(2 \delta D_k)$ and $d_k = (\delta D_k - L_k)/(2\delta D_k)$.
Note that $b_k + d_k=1$ and if $b_k$ and $d_k$ lie outside $0$ and
1, we reset the minimum value to 0 and the maximum value to 1. Let
$Q^P_k = b_k (p^P_k + p^D_k)$ and $Q^D_k = d_k (p^P_k + p^D_k)$,
then we take the importance sampling weights to be
\begin{eqnarray} \label{eq:rep_2}
q^P_k &=& \left(1- \epsilon \left( \frac{s}{\phi} \right)^\kappa
\right) p^P_k + \epsilon \left( \frac{s}{\phi} \right)^\kappa Q^P_k
\\
q^D_k &=& \left(1- \epsilon \left( \frac{s}{\phi} \right)^\kappa
\right) p^D_k + \epsilon \left( \frac{s}{\phi} \right)^\kappa Q^D_k.
\end{eqnarray}
This results in increasing/decreasing the production and decay rates
of the mRNA of gene $k$ to push the $\mathbf{W}^y (s)$ towards
$\mathbf{y}_i$. As in the Lotka-Volterra model there is an increased
push as $s$ approaches $\tilde{\phi}$. We found that $\epsilon =0.2$
and $\kappa =4$ worked well with typically between 1.3 and 2.0 times
as many simulations typically required if importance sampling was
not used.
%
%

\begin{figure}
\epsfig{file=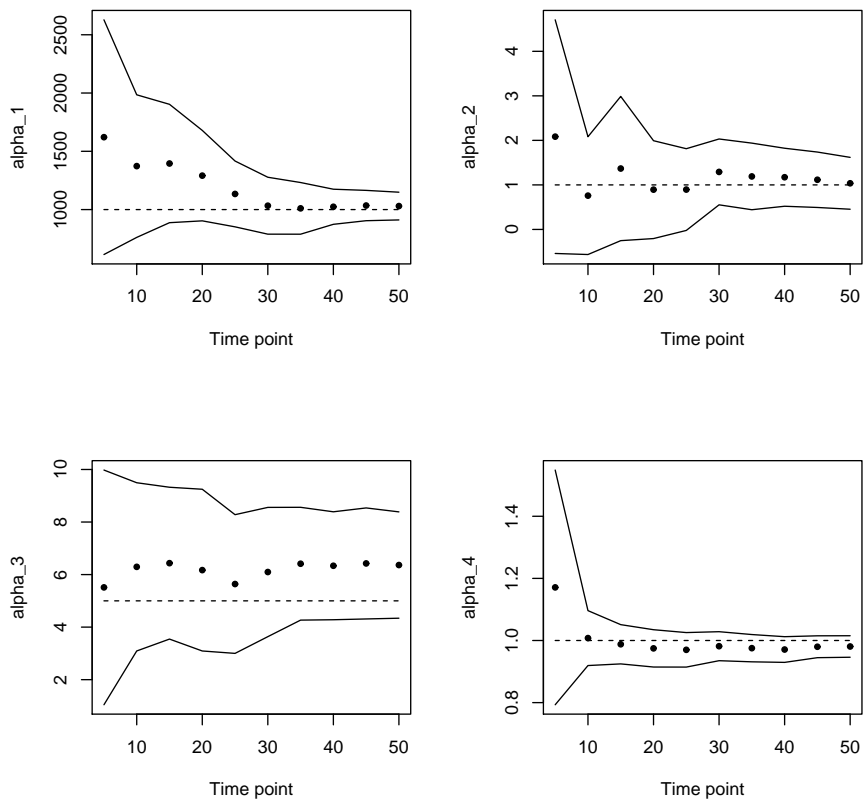, height=12cm, width=12cm}
\caption{Plots of mean of $\alpha$ parameters plus/minus two
standard deviations.} \label{fig:5}
\end{figure}

The total number of simulations across the 50 time points was
583,272,179. There is considerable variation in the number of
simulations per time point ranging from just over a million for time
points 3, 9 and 13 to over 118 million ($20.3\%$ of all simulations)
for time point 41. Time point 41 stood out with no other time point
requiring more than 33 million simulations. In Figure \ref{fig:5},
the estimated posterior means of the $\alpha$ parameters are given
along with lines denoting the mean plus and minus two standard
deviations for every fifth time point from time point 5 to 50. A
similar plot is observed for $\omega$. In all cases the estimated
posterior mean after 50 time points are close to the true simulated
parameters with good estimation of the parameters being apparent
from as few as 10 time points for some parameters.  This suggests
that the mRNA levels are very informative about the parameters of
the model. However, there is greater uncertainty in $\alpha_3$,
which governs the protein production and decay rates, than the other
parameters. This is not surprising as the estimation of $\alpha_3$
depends exclusively on the unobserved protein levels. Similar
observations concerning parameter estimates were seen with other
simulated data sets. Finally, the estimated posterior means and
standard deviations for $\bthet_\alpha | \mathbf{y}_{0:50},
\mathbf{z}_0$ are given in Table \ref{table:2}.


\begin{table}
{Table 3: Estimated posterior means and standard deviations for
$\bthet_\alpha | \mathbf{y}_{0:50}, \mathbf{z}_0$.}
{%
\begin{tabular}{|c|c|c|c|c|c|}
\hline Parameter & $\alpha_1$ &
$\alpha_2$ & $\alpha_3$ & $\alpha_4$ & $\omega$ \\
\hline  Mean &  1030 & 1.036 & 6.363 & 0.9807 & 2.079 \\
\hline St. Dev. & 59.49 &  0.2912 & 1.012 & 0.0173 & 0.0745 \\
\hline
\end{tabular}}
\label{table:2}
\end{table}

\section{Conclusions} \label{S:Conc}

This paper has introduced a sequential Monte Carlo (SMC) algorithm
for discretely observed Markov processes which can successfully and
efficiently obtain samples from the posterior distribution of the
parameters. The two key innovations of coupled simulations and a
simple, yet effective, importance sampler have been central to this
success and are complementary to each other. Both innovations  offer
improvements throughout the SMC algorithm, however, the coupled
simulations are particularly effective in the early stages where
there is greater uncertainty about the parameters with larger sets
$\mathcal{B}$. The importance sampling on the other hand is
particularly useful when the outcome $\mathbf{Y} (t_i)$ is unusual
given $\mathbf{X} (t_{i-1})$ and $\bthet$. The coupled simulations
are straightforward to implement given the reparameterisation,
whereas the importance sampling is problem specific but the
importance sampling approaches taken in this paper, especially the
local linearisation in Section \ref{S:LV}, should be generally
applicable.

There are a few concluding remarks to make about the SMC algorithm.
Firstly, it is trivial to parallelise as at any given time point
simulations can be run independently. Thus as the simulations are
the time consuming part of the SMC algorithm efficient use of
available computing power can be made. Secondly, we have assumed
that the observations from the Markov process are exact, if only
sometimes partial. It is however straightforward to extend the SMC
algorithm to data with observation error.

The SMC algorithm has its origins in the ABC algorithm (\cite{Tav},
\cite{BZB}) and the ideas developed in this paper could be more
widely applied to refining ABC algorithms. The sequential approach
of building up the simulation of a stochastic process with
refinement of the posterior distribution could be widely used. Also
as noted in \cite{WKP}, simulating a stochastic processes in stages
allows for greater precision to be used in the agreement between the
simulated and observed data without severely compromising the
acceptance probability. Moreover, coupled simulations and in
particular, importance sampling within simulations are worth
considering in the implementation of ABC algorithms. Whilst,
considerable attention in the ABC literature has been devoted to
choice of $\bthet$ (for example, MCMC-ABC, \cite{Marjoram} and
SMC-ABC, \cite{SFT}) and the choice and evaluation of summary
statistics (for example, local-linear regression, \cite{BZB} and
semi-automatic ABC, \cite{FearnheadPrangle}), there has been little
research into improvement of the simulation process to make the ABC
algorithm more efficient. As illustrated in this paper it is
possible to improve on the simulation process without significantly
compromising the efficiency of the simulation process.

%
%

%
%



\section*{Acknowledgements}

The author was supported by the  Engineering and Physical Sciences
Research Council under grant EP/J008443/1.

\appendix

\section*{Appendix: Random variable sets}

We outline how sets of values can be drawn from a random variable
and how these can be exploited to give unbiased estimates of key
quantities of interest. In particular, we show how this relates to
the construction and use of $\mathcal{B}$ in Section \ref{ss:smc}.

Let a random variable $X$ have probability density function $f(x)$
and let $\kappa = \sup_x \{ f (x) \}$. Then if $U \sim U[0,1]$, let
$\mathcal{A}_U = \{ x : f(x) \geq U \kappa \}$ be a set drawn from
$X$. For any function $g(\cdot)$,
\begin{eqnarray} \label{eq:app:2} \hat{\theta}
=  \kappa \int_{x \in \mathcal{A}_U} g(x) \, dx \end{eqnarray} is an
unbiased estimate of $\theta = E[g(X)]$, since
\begin{eqnarray} \label{eq:app:3} E[\hat{\theta}]
&=& \int_0^1  \kappa \int_{x \in \mathcal{A}_u} g(x) \, dx \, du
\nonumber \\
&=& \kappa \int_{-\infty}^\infty g(x) \int_0^1 1_{\{ x \in
\mathcal{A}_u \}} \, du \, dx \nonumber \\
&=&\kappa \int_{-\infty}^\infty g(x) \frac{f(x)}{\kappa} \, dx =
\theta.
\end{eqnarray}

Let $\mathcal{B}_u$ denote the set $\mathcal{B}$ constructed in
Section \ref{ss:smc}, step (c), with explicit dependence on $u$.
Note that the maximum of the probability density function of
$N(\tilde{\phi}, \sigma_\phi)$ is  $\kappa = 1/(\sqrt{2 \pi }
\sigma_\phi)$, independent of $\tilde{\phi}$. It then follows from
\eqref{eq:app:3} that $\varpi/(\sqrt{2 \pi } \sigma_\phi)$, given by
\eqref{eq:pi:2} satisfies
\begin{eqnarray} \label{eq:pi:3}
E \left[ \frac{\varpi}{\sqrt{2 \pi } \sigma_\phi} \right] &=& \int
\frac{1}{\sqrt{2 \pi } \sigma_\phi} \int_0^1 \left\{ \int_{\phi \in
\mathcal{B}_u} k_\phi \, d \phi \, du \right\} f_{\mathcal{R}}
(\mathbf{a}; (\tilde{\bthet}^\ast, \phi), \mathbf{x}_{i-1}) \,
d\mathbf{a}
\nonumber \\
&=& \int \int k_\phi \pi (\phi | \tilde{\bthet}^\ast, \mathbf{X}
(t_{i-1})= \mathbf{x}_{i-1}) f_{\mathcal{R}} (\mathbf{a};
(\tilde{\bthet}^\ast, \phi), \mathbf{x}_{i-1}) \, d\mathbf{a} \, d
\phi
\nonumber \\
&=& \pi (\mathbf{Y} (t_i) = \mathbf{y}_i | \tilde{\bthet}^\ast,
\mathbf{X} (t_{i-1}) = \mathbf{x}_{i-1}) \end{eqnarray} as required.

\end{document}